# Extended Jackiw-Pi Model and Its Supersymmetrization

Hitoshi Nishino[1)] and Subhash Rajpoot[2)]

*Department of Physics & Astronomy*
*California State University*
*1250 Bellflower Boulevard*
*Long Beach, CA 90840*

## Abstract

We present an extended version of the so-called Jackiw-Pi (JP) model in three dimensions, and perform its supersymmetrization. Our field content has three multiplets: (i) Yang-Mills vector multiplet $(A_\mu{}^I, \lambda^I)$, (ii) Parity-odd extra vector multiplet $(B_\mu{}^I, \chi^I)$, and (iii) Scalar multiplet $(C^I, \rho^I; f^I)$. The bosonic fields in these multiplets are the same as the original JP-model, except for the auxiliary field $f^I$ which is new, while the fermions $\lambda^I$, $\chi^I$ and $\rho^I$ are their super-partners. The basic difference from the original JP-model is the presence of the kinetic term for $C^I$ with its modified field-strength $H_\mu{}^I \equiv D_\mu C^I + m B_\mu{}^I$. The inclusion of the $C^I$-kinetic term is to comply with the recently-developed tensor hierarchy formulation for supersymmetrization.

---

[1)] E-Mail: hnishino@csulb.edu
[2)] E-Mail: rajpoot@csulb.edu

**1. Introduction.** Ever since the work of Deser-Jackiw-Templeton [1], three dimensional (3D) gauge theory has drawn considerable attention. Their potential applications covers the wide range of fields, such as the condensed matter phenomena, high-$T_c$ superconductivity, and quantum Hall effect. In these lower-dimensional models, the important issue is the mass of gauge fields. For example, in 3D there is a special topological mass term called Chern-Simons (CS) term that preserves the original gauge symmetry.

However, the drawback with the CS topological mass term is the loss of parity-invariance, due to the presence of the $\epsilon^{\mu\nu\rho}$-tensor. To overcome this drawback, Jackiw and Pi have presented a model that preserves the parity by considering two vector fields with opposite parity transformations, generating a mass-gap through Chern-Simons-like term [2].

The consistency of physical states of Jackiw-Pi (JP) model [2] was studied in the Hamiltonian approach [3], and new symmetries with gauge-fixing were discovered [4] in the BRS formulation. Based on the Bonora-Tonin superfield formalism [5], BRS-symmetry of JP-model [2] was analyzed in [6]. The algebraic method of quantization was presented in [7]. The key ingredients for quantization, such as BRS invariance, gauge-fixing, and Slavnov-Taylor identity were studied in [8]. In 3D Schouten-ghost-free gravity, in the Hamiltonian formalism, Deser, Ertl and Grumillier [9] have demonstrated the bifurcation effect, namely, the clash between two local invari-ances. It is conjectured that such a bifurcation effect could appear in the JP-model, since it conforms two local invariances.

The importance of JP-model can be found in a different context. It has been conjectured that the super-algebra $OSp(1|32)$ is the full symmetry group of M-theory [10][11]. It was pointed out in [12] that CS theory for the super-algebra $OSp(32|1)$ appears to contain the so-called M-theory matrix models [13]. Therefore the aforementioned advantage of JP-model over CS theory mandates the supersymmetrization of the original JP-model [2].

The original JP-model [2] has the following lagrangian in our notation:

$$\mathcal{L}_{\rm JP} = -\tfrac{1}{4}(F_{\mu\nu}{}^I)^2 - \tfrac{1}{4}(G_{\mu\nu}{}^I)^2 + \tfrac{1}{2}m\,\epsilon^{\mu\nu\rho}F_{\mu\nu}{}^I B_\rho{}^I \quad, \tag{1.1}$$

where $D_\mu$ is the usual Yang-Mills (YM) gauge-covariant derivative, while $F_{\mu\nu}{}^I$ and $G_{\mu\nu}{}^I$ are the field strengths of $A_\mu{}^I$ and $B_\mu{}^I$ defined by [2]

$$F_{\mu\nu}{}^I \equiv +2\partial_{[\mu}A_{\nu]}{}^I + m f^{IJK} A_\mu{}^J A_\nu{}^K \quad, \tag{1.2a}$$

$$\begin{aligned}G_{\mu\nu}{}^I &\equiv +2D_{[\mu}B_{\nu]}{}^I + f^{IJK} F_{\mu\nu}{}^J C^K \\ &\equiv +(2\partial_{[\mu}B_{\nu]}{}^I + 2m f^{IJK} A_{[\mu}{}^J B_{\nu]}{}^K) + f^{IJK} F_{\mu\nu}{}^J C^K \quad.\end{aligned} \tag{1.2b}$$

The vector $B_\mu{}^I$ has its proper 'gauge' invariance:

$$\delta_\beta B_\mu{}^I = D_\mu \beta^I \quad , \quad \delta_\beta C^K = -m\beta^K \quad . \tag{1.3}$$

The latter transformation combined with the peculiar $F \wedge C$-term in (1.2b) maintains the invariance

$$\delta_\beta G_{\mu\nu}{}^I = 0 \quad . \tag{1.4}$$

After the recent development of non-Abelian tensor formulations [14][15], the sophisticated structures (1.1) through (1.4) can be now understood as a special case of more general 'tensor hierarchy' whose supersymmetrization has been also accomplished. Therefore it is imperative to encompass the JP-model into this newly developed formulation and also study it's supersymmetrization. In passing, we note that the 4D formulation of non-Abelian tensor multiplet [15] has three multiplets: vector multiplet $(A_\mu{}^I, \lambda^I)$, a tensor multiplet $(B_{\mu\nu}{}^I, \chi^I, \varphi)$ and a compensator vector multiplet $(C_\mu{}^I, \rho^I)$. These are 4D multiplets, and their 3D analogs are respectively our present vector multiplet (VM) $(A_\mu{}^I, \lambda^I)$, an extra vector multiplet (EVM) $(B_\mu{}^I, \chi^I)$ and the scalar multiplet (SM) $(C^I, \rho^I)$.[3] The fact that the compensator vector multiplet $(C_\mu{}^I, \rho^I)$ in 4D has its own kinetic term indicates the SM $(C^I, \rho^I)$ in 3D should have also its own kinetic terms to accomplish its supersymmetrization, even though the original JP-model had *no* such a kinetic term for the $C^I$-field [2].

From this viewpoint, we first extend the original JP-model with the kinetic term of the $C^I$-field, and establish its consistency. We refer to this bosonic model as the extended JP-model. Having accomplished this step, we next perform its $N = 1$ supersymmetrization.

In the next section, we present the relevant details of the extended JP-model by including the kinetic term of the $C^I$-field. Subsequently, the super-invariant action is presented in section 3. We investigate the consistency of field equations in section 4. In section 5, we perform superspace re-formulation as an addition confirmation on our component formulation. Concluding remarks are given in section 6.

**2. Extended JP-Model.** As has been alluded to, we comply with the general pattern of tensor-hierarchy formulations [14][15] by introducing the $C^I$-kinetic term[4]

---

[3] We introduce an auxiliary field $f^I$ later for off-shell formulation for the SM.

[4] We assign the engineering dimension 0 (or 1/2) for fundamental bosons (or fermions), so that our lagrangians have the dimension of $(\text{mass})^2$. We can recover the usual $(\text{mass})^4$ for dimensionless action $\widetilde{I}_{\text{JP}} \equiv \kappa^{-2} \int d^3x \, \widetilde{\mathcal{L}}_{\text{JP}}$, by using a constant $\kappa$ with the dimension of length. Accordingly, the gauge-coupling constant $m$ has the dimension of mass.

$$\widetilde{\mathcal{L}}_{\rm JP} = -\tfrac{1}{4}(F_{\mu\nu}{}^I)^2 - \tfrac{1}{4}(G_{\mu\nu}{}^I)^2 - \tfrac{1}{2}(H_\mu{}^I)^2 + \tfrac{1}{2}m\,\epsilon^{\mu\nu\rho}F_{\mu\nu}{}^I B_\rho{}^I \ . \tag{2.1}$$

Here $H_\mu{}^I$ is the $C^I$-field strength [2]

$$H_\mu{}^I \equiv D_\mu C^I + m B_\mu{}^I \ . \tag{2.2}$$

Even though this modified field strength was introduced in the original paper by Jackiw-Pi [2], the kinetic term of the $C$-field was *not explicitly* introduced. As has been mentioned, this modification is motivated by the recently-developed 'tensor hierarchy' formulation [14][15], as a special case. Due to the modified field-strength for $C^I$, the original $B_\mu{}^I$-field equation in [2] is modified to

$$\frac{\delta \widetilde{I}_{\rm JP}}{\delta B_\mu{}^I} = -D_\nu G^{\mu\nu\,I} + \tfrac{1}{2} m \epsilon^{\mu\nu\rho} F_{\nu\rho}{}^I - m H^{\mu\,I} \doteq 0 \ . \tag{2.3}$$

The important consistency question is

$$0 \stackrel{?}{=} D_\mu \left( \frac{\delta \widetilde{I}_{\rm JP}}{\delta B_\mu{}^I} \right) = -D_\mu D_\nu G^{\mu\nu\,I} + \tfrac{1}{2} m \epsilon^{\mu\nu\rho}\, D_{[\mu} F_{\nu\rho]}{}^I - m D_\mu H^{\mu\,I}$$
$$= -\tfrac{1}{2} m f^{IJK} F_{\mu\nu}{}^J G^{\mu\nu\,K} - m D_\mu H^{\mu\,I} \ . \tag{2.4}$$

Note here that these remaining terms vanish exactly due to the $C^I$-field equation:

$$\frac{\delta \widetilde{I}_{\rm JP}}{\delta C^I} = +D_\mu H^{\mu\,I} + \tfrac{1}{2} f^{IJK} F_{\mu\nu}{}^J G^{\mu\nu\,K} \doteq 0 \ . \tag{2.5}$$

In other words, (2.4) is re-casted into

$$0 \stackrel{?}{=} D_\mu \left( \frac{\delta \widetilde{I}_{\rm JP}}{\delta B_\mu{}^I} \right) \equiv -m \left( \frac{\delta \widetilde{I}_{\rm JP}}{\delta C^I} \right) \doteq 0 \quad (Q.E.D.) \tag{2.6}$$

The second equality here is only an identity, similar to the Bianchi identity.

Eq. (2.6) is also related to the invariance of our action $\widetilde{I}_{\rm JP}$ under the vectorial symmetry $\delta_\beta B_\mu{}^I$ in (1.3), $\delta_\beta A_\mu{}^I = 0$, and

$$\delta_\beta (F_{\mu\nu}{}^I, G_{\mu\nu}{}^I, H_\mu{}^I) = (0, 0, 0) \ . \tag{2.7}$$

Because of this property, it is straightforward to confirm $\delta_\beta \widetilde{I}_{\rm JP} = 0$. This action invariance leads to

$$\delta_\beta \widetilde{I}_{\rm JP} = (\delta_\beta B_\mu{}^I) \left( \frac{\delta \widetilde{I}_{\rm JP}}{\delta B_\mu{}^I} \right) + (\delta_\beta C^I) \left( \frac{\delta \widetilde{I}_{\rm JP}}{\delta C^I} \right) = -\beta^I \left[ D_\mu \left( \frac{\delta \widetilde{I}_{\rm JP}}{\delta B_\mu{}^I} \right) + m \left( \frac{\delta \widetilde{I}_{\rm JP}}{\delta C^I} \right) \right] = 0 \ , \tag{2.8}$$

re-producing the previous result (2.6).

There is an alternative better method of variations for supersymmetric variations which we present later. We can show that the general variations of $G$ and $H$-field strengths are

$$\delta G_{\mu\nu}{}^I = +2D_{[\mu}(\widetilde{\delta} B_{\nu]}{}^I) + 2f^{IJK}(\delta A_{[\mu}{}^J)H_{\nu]}{}^K - f^{IJK}(\delta C^J)F_{\mu\nu}{}^K \quad , \tag{2.9a}$$

$$\delta H_\mu{}^I = +D_\mu(\delta C^I) + m(\widetilde{\delta} B_\mu{}^I) \qquad (\widetilde{\delta} B_\mu{}^I \equiv \delta B_\mu{}^I - f^{IJK}C^J \delta A_\mu{}^K) \quad . \tag{2.9b}$$

According to (2.7), the first three terms of (2.1) are manifestly invariant, while the $mB \wedge F$-term yields

$$\delta \left( \tfrac{1}{2} m \, \epsilon^{\mu\nu\rho} B_\mu{}^I F_{\nu\rho}{}^I \right) = + \tfrac{1}{2} m \, \epsilon^{\mu\nu\rho} (\widetilde{\delta} B_\mu{}^I) \, F_{\nu\rho}{}^I + \tfrac{1}{2} m \, \epsilon^{\mu\nu\rho} (\delta A_\mu{}^I) \, G_{\nu\rho}{}^I \quad . \tag{2.10}$$

In other words, neither the *bare* $B$ nor the *bare* $C$-field term arise in terms of the modified variation $\widetilde{\delta} B_\mu{}^I$, so that the invariance $\delta_\beta \widetilde{I}_{\rm JP} = 0$ becomes manifest.

**3. N = 1 Superinvariant Action.** As has been mentioned, for supersymmetrization of the extended JP-model, we introduce the three multiplets: (i) VM $(A_\mu{}^I, \lambda^I)$, (ii) EVM $(B_\mu{}^I, \chi^I)$, and (iii) SM $(C^I, \rho^I; f^I)$, where $f^I$ is an auxiliary field, such that all of our multiplets are *off shell*. Our total action $I \equiv \kappa^{-2} \int d^3x \, \mathcal{L}$ has the lagrangian

$$\begin{aligned}
\mathcal{L} = & -\tfrac{1}{4}(F_{\mu\nu}{}^I)^2 + \tfrac{1}{2}(\overline{\lambda}^I \slashed{D} \lambda^I) - \tfrac{1}{4}(G_{\mu\nu}{}^I)^2 + \tfrac{1}{2}(\overline{\chi}^I \slashed{D} \chi^I) - \tfrac{1}{2}(H_\mu{}^I)^2 + \tfrac{1}{2}(\overline{\rho}^I \slashed{D} \rho^I) \\
& + \tfrac{1}{2} m \, \epsilon^{\mu\nu\rho} B_\mu{}^I F_{\nu\rho}{}^I + m(\overline{\lambda}^I \chi^I) + m(\overline{\chi}^I \rho^I) + \tfrac{1}{2}(f^I)^2 - \tfrac{1}{4} f^{IJK}(\overline{\chi}^I \gamma^{\mu\nu} \rho^J) F_{\mu\nu}{}^K \\
& - \tfrac{1}{2} f^{IJK}(\overline{\lambda}^I \gamma^\mu \chi^J) H_\mu{}^K + \tfrac{1}{4} f^{IJK}(\overline{\lambda}^I \gamma^{\mu\nu} \rho^J) G_{\mu\nu}{}^K + \tfrac{1}{4} h^{IJ,KL}(\overline{\lambda}^I \lambda^K)(\overline{\rho}^J \rho^L) \\
& - \tfrac{1}{32} h^{IJ,KL}(\overline{\lambda}^I \gamma_\mu \lambda^J)(\overline{\chi}^K \gamma^\mu \chi^L) + \tfrac{1}{16} h^{IJ,KL}(\overline{\lambda}^I \lambda^K)(\overline{\chi}^J \chi^L) \quad ,
\end{aligned} \tag{3.1}$$

where $h^{IJ,KL} \equiv f^{IJM} f^{MKL}$. The definition of the field strengths $F$ and $G$ are exactly the same as (1.2), while that of $H$ is given by (2.2). These field strengths satisfy their proper Bianchi identities:

$$D_{[\mu} F_{\nu\rho]}{}^I \equiv 0 \quad , \qquad D_{[\mu} G_{\nu\rho]}{}^I \equiv + f^{IJK} F_{[\mu\nu}{}^J H_{\rho]}{}^K \quad , \qquad D_{[\mu} H_{\nu]}{}^I \equiv + \tfrac{1}{2} m G_{\mu\nu}{}^I \quad . \tag{3.2}$$

Similar to the previous section, the invariance $\delta_\beta I = 0$ under $\beta$-transformation is easily confirmed.

Our action $I$ is also invariant under $N = 1$ supersymmetry

$$\delta_Q A_\mu{}^I = +(\overline{\epsilon} \gamma_\mu \lambda^I) \quad , \tag{3.3a}$$

$$\delta_Q \lambda^I = +\tfrac{1}{2}(\gamma^{\mu\nu} \epsilon) F_{\mu\nu}{}^I \quad , \tag{3.3b}$$

$$\delta_Q B_\mu{}^I = +(\overline{\epsilon}\gamma_\mu\chi^I) - f^{IJK}(\overline{\epsilon}\gamma_\mu\lambda^J)C^K \quad , \quad \widetilde{\delta}_Q B_\mu{}^I = +(\overline{\epsilon}\gamma_\mu\chi^I) \quad , \tag{3.3c}$$

$$\delta_Q \chi^I = + \tfrac{1}{2}(\gamma^{\mu\nu}\epsilon)G_{\mu\nu}{}^I - \tfrac{1}{2} f^{IJK}\left[\epsilon(\overline{\lambda}^J\rho^K) - (\gamma_\mu\epsilon)(\overline{\lambda}^J\gamma^\mu\rho^K)\right] \quad , \tag{3.3d}$$

$$\delta_Q C^I = +(\overline{\epsilon}\rho^I) \quad , \tag{3.3e}$$

$$\delta_Q \rho^I = -(\gamma^\mu\epsilon)H_\mu{}^I - \epsilon f^I - \tfrac{1}{2} f^{IJK}\epsilon(\overline{\lambda}^J\chi^K) \quad , \tag{3.3f}$$

$$\delta_Q f^I = +(\overline{\epsilon}\slashed{D}\rho^I) + m(\overline{\epsilon}\chi^I) - \tfrac{1}{4} f^{IJK}(\overline{\epsilon}\gamma^{\mu\nu}\chi^J)F_{\mu\nu}{}^K + \tfrac{1}{4} f^{IJK}(\overline{\epsilon}\gamma^{\mu\nu}\lambda^J)G_{\mu\nu}{}^K$$
$$+ \tfrac{1}{2} h^{IJ,KL}(\overline{\epsilon}\rho^K)(\overline{\lambda}^J\lambda^L) \equiv \overline{\epsilon}\left(\frac{\delta I}{\delta \overline{\rho}^I}\right) \quad . \tag{3.3g}$$

Notice that there is *no* fermionic-quadratic terms in $\delta_Q\lambda$, while $\lambda\rho$ or $\lambda\chi$-terms exist in $\delta\chi$ and $\delta\rho$, respectively. They are determined by the supersymmetric invariance $\delta_Q I$ at $\mathcal{O}(m\Phi^3)$ or $\mathcal{O}(m^0\Phi^4)$, where the symbol $\Phi$ stands for any fundamental field in our system, which may contain derivative(s). Our multiplets VM and EVM are all *off-shell*, as can readily be established by counting their degrees of freedom (DOF) $1+1$ (on-shell), and $2+2$ (on-shell). Our SM has $1+1$ (on-shell) and $2+2$ (off-shell) DOF, because the auxiliary field $f^I$ carries one off-shell DOF. The $C^I$-field plays the role of Nabmu-Goldstone field that is absorbed into the longitudinal component of $B_\mu{}^I$, making the latter massive. For completeness, the DOF of our fields are listed in Table 1 below.

| DOF before Absorptions | $A_\mu{}^I$ | $\lambda^I$ | $B_\mu{}^I$ | $\chi^I$ | $C^I$ | $\rho^I$ | $f^I$ |
|---|---|---|---|---|---|---|---|
| Physical | 1 | 1 | 1 | 1 | 1 | 1 | 0 |
| Unphysical & Physical | 2 | 2 | 2 | 2 | 1 | 2 | 1 |

| DOF after Absorptions | $A_\mu{}^I$ | $\lambda^I$ | $B_\mu{}^I$ | $\chi^I$ | $C^I$ | $\rho^I$ | $f^I$ |
|---|---|---|---|---|---|---|---|
| Physical | 1 | 1 | 2 | 2 | 0 | 0 | 0 |
| Unphysical & Physical | 2 | 2 | 3 | 4 | 0 | 0 | 1 |

Table 1: DOF of Our Field Content

In the unphysical & physical DOF after absorptions for the EVM and SM, the $\chi$ and $\rho$-fields form a Dirac fermion with 4 off-shell DOF.

The invariance confirmation $\delta_Q I = 0$ is summarized as follows. They are confirmed order-by-order in terms of the power of fundamental fields, such as $\Phi^2$, $\Phi^3$, $\cdots$. First, at the quadratic order, there are two categories of terms: (I) $m^0\Phi^2$-terms and (II) $m\Phi^2$-terms. The sector (I) is rather a routine confirmation, while there is one subtlety in sector (II), associated with the variation of the $mF\wedge B$-term in the lagrangian. This is because $\delta_Q B_\mu{}^I$ in the first expression in (3.3c) contains the *bare* $C$-field. However, as the arbitrary variation of the $mF\wedge B$-term shows in (2.10), the *bare* $C$-field term does *not* arise.

Relevantly, the supersymmetry transformation rule $\widetilde{\delta}_Q B_\mu{}^I$ is the second expression in (3.3c). This is a common feature of a potential field whose field strength is a modified (generalized) CS-term.

Second, the cubic-order terms are type (I) $m^0\Phi^3$-terms and type (II) $m\Phi^3$-terms. For the former, there are eight sectors (i) $\chi FH$, (ii) $\rho FG$, (iii) $\lambda GH$, (iv) $\lambda\chi D\rho$, $\chi\rho D\chi$, or $\rho\lambda D\chi$, (v) $\chi fF$, (vi) $\lambda fG$, (vii) $\chi fF$, and (viii) $\lambda fG$. The key relationships needed are the Bianchi identities (3.2). The type (II) $m\Phi^3$-terms have four sectors: (i) $m\lambda\rho^2$, (ii) $m\lambda\chi^2$, (iii) $m\rho\lambda^2$ and (iv) $m\rho\chi^2$. The subtlety here is that some quadratic-fermion terms in $\delta_Q\lambda$, $\delta_Q\chi$ and $\chi_Q\rho$ are all involved in these sectors, due to the existence of $m(\text{Fermion})^2$-terms in the lagrangian.

Third, the quartic terms are of the type $m^0\Phi^4$, and there are seven sectors: (i) $\chi^2\lambda F$, (ii) $\lambda^2\chi G$, (iii) $\chi^2\rho H$, (iv) $\rho^2\chi G$, (v) $\lambda^2\rho H$, (vi) $\rho^2\lambda F$, and (vii) $\rho\lambda^2 f$. These determine the quadratic-fermion terms in $\delta_Q\lambda$, $\delta_Q\chi$ and $\delta_Q\rho$, and quartic-fermion terms in the lagrangian. After tedious cancellations and by the use of the relationships, such as the Jacobi identity $h^{[IJ,K]L} \equiv 0$, the final form of the lagrangian is obtained, *e.g.*, the absence of the $\chi^2\rho^2$-terms in the lagrangian, and the absence of (Fermion)$^2$-terms in $\delta_Q\lambda$. We have found that these structures are uniquely determined by the cancellation of these terms at $m^0\Phi^4$. The $f^I$-dependent terms cancel each other, justifying the $\rho\lambda^2$-term in $\delta_Q f^I$ and $f^I$-linear term in $\delta_Q\rho^I$. As for all of the auxiliary-field $f^I$-dependent terms in $\delta_Q I$, they cancel themselves manifestly, if we use the last expression of (3.3g).

As is the common feature of non-Abelian tensor theories [14][15] (or extra vector as its special case), our lagrangian (3.1) has terms that are *not*-renormalizable. This is established as follows. In 3D, the most conventional physical dimension for a boson (or a fermion) is $1/2$ (or 1),[5] so that the gauge-coupling constant has dimension 0. Therefore, the cubic terms, *e.g.*, $f^{IJK}(\overline{\chi}^I\gamma^{\mu\nu}\rho^J)F_{\mu\nu}{}^K$ with the dimension $1+1+3/2 = 7/2 > 3$, or the quartic terms, *e.g.*, $h^{IJ,KL}(\overline{\lambda}^I\lambda^K)(\overline{\rho}^J\rho^L)$ with the dimension $1 \times 4 = 4 > 3$ are *not* renormalizable.

However, we expect that the renormalizability of the supersymmetric JP-model presented here will be much improved from its original form due to supersymmetry, a feature common to all supersymmetric theories. Typical examples are non-linear sigma-models, which are originally not renormalizable, but become even *finite* by supersymmetrization, such as *finite* $N=2$ supersymmetric sigma-models [16].

---

[5] These conventional dimensions are different from our engineering dimensions: $d=0$ (or $d=1/2$) for bosons (or fermions).

**4. Consistency of Field Equations.** We first list up the field equations of all of our fields obtained from our action $I$ of (3.1):

$$\frac{\delta I}{\delta \overline{\lambda}^I} = + \not{D}\lambda^I + m\chi^I - f^{IJK}(\gamma^\mu \chi^J) H_\mu{}^K + \tfrac{1}{4} f^{IJK}(\gamma^{\mu\nu} \rho^J) G_{\mu\nu}{}^K + \tfrac{1}{2} h^{IJ,KL} \lambda^K (\overline{\rho}^J \rho^L)$$
$$- \tfrac{1}{16} h^{IJ,KL}(\gamma_\mu \lambda^J)(\overline{\chi}^K \gamma^\mu \chi^L) + \tfrac{1}{8} h^{IJ,KL} \lambda^K (\overline{\chi}^J \chi^L) \doteq 0 \;, \tag{4.1a}$$

$$\frac{\delta I}{\delta \overline{\chi}^I} = + \not{D}\chi^I + m\lambda^I + m\rho^I - \tfrac{1}{4} f^{IJK}(\gamma^{\mu\nu} \rho^J) F_{\mu\nu}{}^K - \tfrac{1}{2} f^{IJK}(\gamma^\mu \lambda^J) H_\mu{}^K$$
$$- \tfrac{1}{16} h^{IJ,KL}(\gamma_\mu \chi^J)(\overline{\lambda}^K \gamma^\mu \lambda^L) + \tfrac{1}{8} h^{IJ,KL} \chi^K (\overline{\lambda}^J \lambda^L) \doteq 0 \;, \tag{4.1b}$$

$$\frac{\delta I}{\delta \overline{\rho}^I} = + \not{D}\rho^I + m\chi^I - \tfrac{1}{4} f^{IJK}(\gamma^{\mu\nu} \chi^J) F_{\mu\nu}{}^K + \tfrac{1}{4} f^{IJK}(\gamma^{\mu\nu} \lambda^J) G_{\mu\nu}{}^K$$
$$+ \tfrac{1}{2} h^{IJ,KL} \rho^K (\overline{\lambda}^J \lambda^L) \doteq 0 \;, \tag{4.1c}$$

$$\frac{\delta I}{\delta A_\mu{}^I} = - D_\nu F^{\mu\nu\,I} + \tfrac{1}{2} m \epsilon^{\mu\nu\rho} G_{\nu\rho}{}^I - \tfrac{1}{2} m f^{IJK} \left[ (\overline{\lambda}^J \gamma^\mu \lambda^K) + (\overline{\chi}^J \gamma^\mu \chi^K) + (\overline{\rho}^J \gamma^\mu \rho^K) \right]$$
$$+ f^{IJK} G^{\mu\nu\,J} H_\nu{}^K + f^{IJK} C^J \left( \frac{\delta I}{\delta B_\mu{}^K} \right) - \tfrac{1}{2} f^{IJK} D_\nu (\overline{\chi}^J \gamma^{\mu\nu} \rho^K)$$
$$+ \tfrac{1}{2} h^{IJ,KL} (\overline{\lambda}^K \gamma^{\mu\nu} \rho^L) H_\nu{}^J \doteq 0 \;, \tag{4.1d}$$

$$\frac{\delta I}{\delta B_\mu{}^I} = - D_\nu G^{\mu\nu\,I} + \tfrac{1}{2} m \epsilon^{\mu\nu\rho} F_{\nu\rho}{}^I - m H^{\mu\,I}$$
$$- \tfrac{1}{2} m f^{IJK} (\overline{\lambda}^J \gamma^\mu \chi^K) - \tfrac{1}{2} m f^{IJK} D_\nu (\overline{\lambda}^J \gamma^{\mu\nu} \rho^K) \doteq 0 \;, \tag{4.1e}$$

$$\frac{\delta I}{\delta C^I} = + D_\mu H^{\mu\,I} + \tfrac{1}{2} f^{IJK} F_{\mu\nu}{}^J G^{\mu\nu\,K} - \tfrac{1}{2} m f^{IJK} (\overline{\lambda}^J \rho^K)$$
$$- \tfrac{1}{8} h^{IJ,KL} (\overline{\chi}^J \gamma^{\mu\nu} \rho^K) G_{\mu\nu}{}^L + \tfrac{1}{4} h^{IJ,KL} \left[ (\overline{\lambda}^J \gamma^\mu \lambda^K) + (\overline{\chi}^J \gamma^\mu \chi^K) \right] H_\mu{}^L$$
$$+ \tfrac{1}{8} h^{IJ,KL} (\overline{\lambda}^J \gamma^{\mu\nu} \rho^K) F_{\mu\nu}{}^L - \tfrac{1}{4} h^{IJ,KL} (\overline{\lambda}^K \gamma^{\mu\nu} \rho^L) F_{\mu\nu}{}^J$$
$$+ \tfrac{1}{2} f^{IJK} \left( \overline{\lambda}^J \frac{\delta I}{\delta \overline{\chi}^K} \right) + \tfrac{1}{2} f^{IJK} \left( \overline{\chi}^J \frac{\delta I}{\delta \overline{\lambda}^K} \right) \doteq 0 \;, \tag{4.1f}$$

$$\frac{\delta I}{\delta f^I} = + f^I \doteq 0 \;. \tag{4.1g}$$

As has been discussed in the non-supersymmetric case with (2.6), the most crucial consistency question is whether the divergence of the $B_\mu{}^I$-field equation vanishes. This is confirmed as the supersymmetric generalization of the purely bosonic case. The result is simply

$$0 \stackrel{?}{=} D_\mu \left( \frac{\delta I}{\delta B_\mu{}^I} \right) \equiv -m \left( \frac{\delta I}{\delta C^I} \right) \doteq 0 \;. \tag{4.2}$$

Note that the middle equality here is an identity, and *no* field equation has been used. This is formally the same as the non-supersymmetric case (2.6), since this is nothing but the $\delta_\beta$-invariance of our action:

$$\delta_\beta I = -\beta^I \left[ D_\mu \left( \frac{\delta I}{\delta B_\mu{}^I} \right) + m \left( \frac{\delta I}{\delta C^I} \right) \right] \equiv 0 \ . \tag{4.3}$$

Note that the second equality in (4.2) can be explicitly confirmed for our field equations (4.1). In particular, when we apply the covariant derivative to (4.1e), all terms cancel themselves, including the quartic-fermion terms. Crucial cancellations occur where identities are needed, such as

$$(k^{JK,I,LM} + k^{LM,I,JK})(\overline{\lambda}^J \gamma_\mu \lambda^K)(\overline{\chi}^L \gamma^\mu \chi^M) \equiv 0 \ , \tag{4.4a}$$

$$k^{JK,I,LM}(\overline{\lambda}^J \lambda^L)(\overline{\chi}^K \chi^M) \equiv 0 \ , \tag{4.4b}$$

where $k^{IJ,K,LM} \equiv f^{IJN} f^{NKP} f^{PLM}$. These identities are confirmed by the relationships

$$k^{[IJ,K],LM} \equiv k^{IJ,[K,LM]} \equiv 0 \ , \quad k^{IJ,K,LM} = -k^{LM,K,IJ} \ . \tag{4.5}$$

We can also confirm similar consistency for the $A_\mu{}^I$-field equation:

$$0 \overset{?}{=} D_\mu \left( \frac{\delta I}{\delta A_\mu{}^I} \right) = -mf^{IJK}\overline{\lambda}^J \left( \frac{\delta I}{\delta \overline{\lambda}^K} \right) - mf^{IJK}\overline{\chi}^J \left( \frac{\delta I}{\delta \overline{\chi}^K} \right) - mf^{IJK}\overline{\rho}^J \left( \frac{\delta I}{\delta \overline{\rho}^K} \right)$$

$$- f^{IJK} H^J \left( \frac{\delta I}{\delta B_\mu{}^K} \right) + f^{IJK} D_\mu \left[ C^J \left( \frac{\delta I}{\delta B_\mu{}^K} \right) \right] \doteq 0 \ . \tag{4.6}$$

This is nothing but the YM-gauge invariance

$$\delta_\alpha A_\mu{}^I = D_\mu \alpha^I \ , \quad \delta_\alpha(B_\mu{}^I, \ C^I, \ \lambda^I, \ \chi^I, \ \rho^I) = -mf^{IJK}\alpha^J(B_\mu{}^K, \ C^K, \ \lambda^K, \ \chi^K, \ \rho^K) \tag{4.7}$$

of our action:

$$\delta_\alpha I = + (\delta_\alpha A_\mu{}^I) \left( \frac{\delta I}{\delta A_\mu{}^I} \right) + (\delta_\alpha B_\mu{}^I) \left( \frac{\delta I}{\delta B_\mu{}^I} \right) + (\delta_\alpha C^I) \left( \frac{\delta I}{\delta C^I} \right)$$

$$+ (\delta_\alpha \overline{\lambda}^I) \left( \frac{\delta I}{\delta \overline{\lambda}^I} \right) + (\delta_\alpha \overline{\chi}^I) \left( \frac{\delta I}{\delta \overline{\chi}^I} \right) + (\delta_\alpha \overline{\rho}^I) \left( \frac{\delta I}{\delta \overline{\rho}^I} \right) \tag{4.8a}$$

$$= -\alpha^I D_\mu \left( \frac{\delta I}{\delta A_\mu{}^I} \right) - mf^{IJK}\alpha^J \left[ \overline{\lambda}^K \left( \frac{\delta I}{\delta \overline{\lambda}^I} \right) + \overline{\chi}^K \left( \frac{\delta I}{\delta \overline{\chi}^I} \right) + \overline{\rho}^K \left( \frac{\delta I}{\delta \overline{\rho}^I} \right) \right]$$

$$- f^{IJK} \alpha^I H^J \left( \frac{\delta I}{\delta B_\mu{}^K} \right) + f^{IJK} \alpha^I D_\mu \left[ C^J \left( \frac{\delta I}{\delta B_\mu{}^K} \right) \right] \equiv 0 \ . \tag{4.8b}$$

By the use of (4.3), the $(\delta_\alpha C)(\delta I/\delta C)$-term in (4.8a) is replaced by $m^{-1}\alpha\, CD(\delta I/\delta B)$-term, which in turn is replaced by

$$f^{IJK}\alpha^J C^K D_\mu \left(\frac{\delta I}{\delta B_\mu{}^I}\right) = f^{IJK}\alpha^I \left\{ D_\mu \left[ C^J \left(\frac{\delta I}{\delta B_\mu{}^K}\right)\right] - (H_\mu{}^J - mB_\mu{}^J)\left(\frac{\delta I}{\delta B_\mu{}^K}\right)\right\} , \quad (4.9)$$

and the last $mB(\delta I/\delta B)$-term will be cancelled by the like-term in (4.8a). Eventually, we end up with (4.8b).

**5. Superspace Re-Formulation.** We can reconfirm our component-field result in terms of superspace language [17] The basic ingredients are the superfield strengths $F_{AB}{}^I$, $G_{AB}{}^I$ and $H_A{}^I$,[6] satisfying the Bianchi identities

$$+\frac{1}{2}\nabla_{[A}F_{BC)}{}^I - \frac{1}{2}T_{[AB|}{}^D F_{D|C)}{}^I \equiv 0 , \quad (5.1a)$$

$$+\frac{1}{2}\nabla_{[A}G_{BC)}{}^I - \frac{1}{2}T_{[AB|}{}^D G_{D|C)}{}^I - \frac{1}{2}f^{IJK}F_{[AB|}{}^J H_{|C)}{}^K \equiv 0 , \quad (5.1b)$$

$$+\nabla_{[A}H_{B)}{}^I - T_{AB}{}^C H_C{}^I - mG_{AB}{}^I \equiv 0 . \quad (5.1b)$$

The constraints at engineering dimensions $0 \le d \le 1$ are

$$T_{\alpha\beta}{}^c = +2(\gamma^c)_{\alpha\beta} , \quad F_{\alpha b}{}^I = -(\gamma_b \lambda^I)_\alpha , \quad G_{\alpha b}{}^I = -(\gamma_b \chi^I)_\alpha , \quad H_\alpha{}^I = -\rho_\alpha{}^I , \quad (5.2a)$$

$$\nabla_\alpha \lambda_\beta{}^I = +\tfrac{1}{2}(\gamma^{cd})_{\alpha\beta} F_{cd}{}^I + C_{\alpha\beta}f^I , \quad (5.2b)$$

$$\nabla_\alpha \chi_\beta{}^I = +\tfrac{1}{2}(\gamma^{cd})_{\alpha\beta} G_{cd}{}^I + \tfrac{1}{2}f^{IJK}C_{\alpha\beta}(\overline{\lambda}^J \rho^K) - \tfrac{1}{2}f^{IJK}(\gamma_c)_{\alpha\beta}(\overline{\lambda}^J \gamma^c \rho^K) , \quad (5.2c)$$

$$\nabla_\alpha \rho_\beta{}^I = -(\gamma^c)_{\alpha\beta} H_c{}^I + \tfrac{1}{2}C_{\alpha\beta}(\overline{\lambda}^J \chi^K) + C_{\alpha\beta}f^I , \quad (5.2d)$$

Other independent components, such as $F_{\alpha\beta}{}^I$ are all zero. The constraints at $d = 3/2$ are

$$\nabla_\alpha f^I = -(\slashed{\nabla}\rho^I)_\alpha - m\chi_\alpha + \tfrac{1}{4}f^{IJK}(\gamma^{bc}\chi^J)_\alpha F_{bc}{}^K - \tfrac{1}{4}(\gamma^{bc}\lambda^J)_\alpha G_{bc}{}^K$$
$$- \tfrac{1}{2}h^{IJ,KL}\rho_\alpha{}^K(\overline{\lambda}^J\chi^L) , \quad (5.3a)$$

$$\nabla_\alpha F_{bc}{}^I = +(\gamma_{[b}\nabla_{c]}\lambda^I)_\alpha , \quad (5.3b)$$

$$\nabla_\alpha G_{bc}{}^I = +(\gamma_{[b}\nabla_{c]}\chi^I)_\alpha - f^{IJK}(\gamma_{[b|}\lambda^J)_\alpha H_{|c]}{}^K + f^{IJK}\rho_\alpha{}^J F_{bc}{}^K , \quad (5.3c)$$

$$\nabla_\alpha H_b{}^J = -\nabla_b \rho_\alpha{}^I - m(\gamma_b \chi^I)_\alpha . \quad (5.3d)$$

The $\rho_\alpha{}^I$-field equation is obtained by the 'on-shell-ness' requirement $f^I \doteq 0$, as usual in off-shell formulation with auxiliary fields. The resulting $\rho_\alpha{}^I$-field equation is consistent with (4.1c) in component which is skipped here. As for $\lambda^I$ and $\chi^I$-field equations, they can be obtained only by the action invariance. We can confirm their consistency with supersymmetry by taking their spinorial derivative $\nabla_\alpha$, yielding the bosonic field equations (4.1d) through (4.1g).

---

[6] We use the superspace indices $A, B, \cdots = (a,\alpha), (b,\beta), \cdots$ for bosonic $a, b, \cdots = 0, 1, 2$ and fermionic $\alpha, \beta, \cdots = 1, 2$ coordinates. Our antisymmetrization in superspace is such as $M_{[AB)} \equiv M_{AB} - (-1)^{AB}M_{BA}$, etc.

Note that the off-shell structure of our system is consistent with our own component result. This also provides the supporting evidence of the total consistency of our system. From this viewpoint, we regard our system is the unique supersymmetrization of the original JP-model [2], which necessitates the existence of the *physical* SM $(C^I, \rho_\alpha{}^I; f^I)$.

## 6. Concluding Remarks.

In this Letter, we have accomplished the $N=1$ off-shell supersymmetrization of the extended JP-model [2]. This necessitates the introduction of the kinetic term of the $C^I$-field.

There are two reasons for our introduction of the kinetic term of the $C^I$-field: First, it is motivated by the recent development of tensor hierarchy formulation [14][15]. The consistency of the $B_\mu{}^I$-field equation is associated with the $\delta_\beta$-invariance of our action which is *not* well stressed in the original JP-model [2]. Second, it excludes the extra constraint $f^{IJK} F_{\mu\nu}{}^J G^{\mu\nu K} \doteq 0$, because this served as the obstruction to supersymmetrizations.

We have also confirmed the total consistency of our supersymmetric system. We have confirmed the identities (4.2) and (4.6) by using our field equations in (4.1). In particular, these consistencies have been explicitly confirmed even with non-trivial fermionic quartic terms. Involving all field equations, this non-trivial confirmation procedure has established the total consistency of our system. Additional confirmation has been performed also in superspace.

Our supersymmetric system is non-trivial. We can *not* simply truncate the kinetic term of the SM $(C^I, \rho^I; f^I)$, because the action invariance no longer respects invariance for the truncated system. This again justifies the necessity of the kinetic terms for $C^I$ and $\rho^I$.

We believe our present result should help in generating other and new consistent topological massive non-Abelian gauge theories and their supersymmetrization.